\documentstyle[12pt]{article}

\title{The structure of neutron deficient Sn isotopes}
\author{A. Holt, T. Engeland, M. Hjorth-Jensen and E. Osnes\\
Department of Physics,
University of Oslo, N-0316 Oslo, Norway}

\begin{document}

\maketitle

\begin{abstract}
In this work we present a realistic shell model calculation
of the low-lying spectra for Sn isotopes with
mass number  $A=102$ --  $108$.
The effective shell model interaction was derived  from a realistic
nucleon-nucleon interaction employing perturbative many-body
techniques.
\end{abstract}

\section{Introduction}

During the last few years through the radioactive nuclear
beams program a rich
variety of data has become available for nuclei far from the stability line.
Recently, substantial progress has been made in the spectroscopic approach
to the neutron deficient doubly magic $^{100}$Sn core. Experimental
spectroscopic data are presently available down to $^{104}$Sn
[1-4].
Thus theoretical shell model calculations for these nuclei
are greatly needed. The starting point for such an analysis is the doubly
magic core $^{100}$Sn. This system has been studied theoretically, using
either approaches inspired by the relativistic Serot-Walecka \cite{sw86}
model, such as the calculations of Hirata {\em et al.} \cite{hir91}
and Nikolaus {\em et al.} \cite{nhm92}, or non-relativistic
models used by Leander {\em et al.} \cite{lean84}. In both approaches it
is concluded that $^{100}$Sn is a reasonably stable closed core.
Thus the Sn isotopes should be well suited for a shell model
analysis.

{}From a theoretical point of view the Sn isotopes are interesting
to study as they are an unique testing ground for nuclear structure
calculations. We have a large sequence of nuclei which can be described in
terms of the neutron degrees of freedom only.
Here, it is often assumed that the seniority or
generalized seniority scheme is a good approximation \cite{tal77}
to the large
shell model problem these isotopes represents.
This can now be tested with our present shell model code
which is able to handle  many of the isotopes.
Also the question  of an additional effective
three-body interaction \cite{mfko83} may be investigated
when the number of valence neutrons becomes significant.

In several papers [1-3] Schubart {\em et al.}
have calculated energy spectra above $^{100}$Sn within the framework
of the spherical
shell model. However, their results have some deficiencies. The two-body
interaction they used had to be scaled by a factor $1.4$ in order to give a
reasonable agreement with experiment, and the E2 transition rates
measured in $^{104}$Sn and $^{106}$Sn are not well understood in
their model.

A realistic shell model calculation for even Sn isotopes with mass number
$A = 102$, $104$ and $106$ was recently published by us in ref.\
\cite{ehho93}.In this work we present a calculation for $^{102-108}$Sn,
including the odd isotopes, using the doubly magic $^{100}$Sn as core.

In the next section we briefly outline the derivation of the effective
interaction and some detailes for the shell model calculation. Our
results are presented in sect. 3, where the energy spectra and also some
aspects of the single particle energies are discussed.

\section{Effective interaction and shell model calculation}

In this work we derive a two-body
neutron-neutron effective interaction
calculated from conventional perturbative many-body techniques
\cite{ko90}.
Here we sketch briefly the main ingredients.

The many-body Schr\"{o}dinger equation for an $A$-nucleon system
is given by
\begin{equation}
                H\Psi_{i}(1,...,A)=E_{i}\Psi_{i}(1,...,A)
                \label{eq:schr}
\end{equation}
with
$H=T+V$,
$T$ being the kinetic energy operator and $V$ the nucleon-nucleon
potential.
For the nucleon-nucleon potential $V$, we adopt here the parameters
of the Bonn A potential
defined in table A.2 of ref.\ \cite{mac89}.
$E_{i}$ and $\Psi_{i}$ are the eigenvalues and eigenfunctions
for a state $i$.
Introducing the auxiliary single-particle potential $U$, $H$ can
be rewritten as
\begin{equation}
          \begin{array}{ccc}H=H_{0}+H_1, \,\,&H_{0}=T+U,
\,\,&H_1=V-U\end{array}.
\end{equation}
If $U$ is chosen such that $H_1$ becomes small, then $H_1$
can be treated as a perturbation.
The eigenfunctions of $H_{0}$
are then the unperturbed wave functions $|\psi _{i} \rangle$ with
$W_{i}$
as the respective eigenvalue.

Usually, one is only interested in solving the Schr\"{o}dinger
equation for certain
low-lying states. It is then customary to divide the Hilbert space
into a model space defined by the operator $P$
\begin{equation}
          P=\sum_{i=1}^{d}|\psi_{i} \rangle \langle \psi_{i} |
\end{equation}
with $d$ being the size of the model space, and an excluded space
defined by the operator $Q$
\begin{equation}
          Q=\sum_{i=d+1}^{\infty}|\psi_{i} \rangle \langle \psi_{i}|
\end{equation}
such that $PQ=0$.
In this work we define the model space to consist of the orbitals in
the $N = 4$ oscillator shell ($1d_{5/2}$, $0g_{7/2}$, $1d_{3/2}$,
$2s_{1/2}$) and the intruder orbital $0h_{11/2}$
from the $N = 5$ oscillator shell.
As we show below, the inclusion of the $h_{11/2}$ orbit to our model space
is needed in order to obtain a quantitative description of the data.

No experimental information is available for the $^{101}$Sn
one-neutron system
to establish the single-particle energies necessary for
the shell model calculation. Thus, these must be evaluated theoretically.
In the present calculation the data are taken from ref.\ \cite{grawe92},
with $
\varepsilon_{d_{3/2}}-\varepsilon_{d_{5/2}} =2.89$ MeV, $%
\varepsilon_{s_{1/2}}-\varepsilon_{d_{5/2}} =2.80$ MeV and $%
\varepsilon_{h_{11/2}}-\varepsilon_{d_{5/2}}$ set to 5.0 MeV.

Eq.\ (\ref{eq:schr}) can be rewritten as
\begin{equation}
PH_{\mathrm{eff}}P\Psi_{i}=P(H_{0}+V_{\mathrm{eff}})
                P\Psi_{i}=E_{i}P\Psi_{i}
\label{eq:eff}
\end{equation}
where $H_{\mathrm{eff}}$  is an effective Hamiltonian acting solely
within the chosen model space.
In this work we have used an iterative approach to the energy-independent
expansion for $H_{\mathrm{eff}}$ advocated by Lee and Suzuki (LS)
\cite{ls80}.The LS expansion for the effective interaction is then formally
given as
\begin{equation}
   H_{\mathrm{eff}}=H_{0}+\lim_{n \rightarrow\infty}R_{n},
\end{equation}
with
\begin{equation}
           R_{n}=\left[1-\hat{Q}_{1}-\sum_{m=2}^{n}\hat{Q}_{m}
                \prod_{k=n-m+1}^{n-1}R_{k}\right]^{-1}\hat{Q}.
\end{equation}
Here we have defined
$\hat{Q}_{m}=\frac{1}{m!}\frac{d^{m}\hat{Q}}{d\Omega^{m}}$, where $\Omega$
is the energy of the interacting nucleons in a medium.
To define the $\hat{Q}$-box, we include all diagrams
through third order in the interaction, as defined in the appendix of
ref.\ \cite{hom92}. The $\hat{Q}$-box consists of all valence-linked,
non-folded and irreducible diagrams, and is conventionally defined in
terms of the $G$-matrix
\begin{equation}
                 G=V+V\frac{Q}{\Omega - QH_0 Q}G.
                 \label{eq:gmat}
\end{equation}
We solve eq.\ (\ref{eq:gmat})
through the so-called
double-partitioning scheme defined in ref.\ \cite{kkko76}. The
operator $Q$ is constructed so as to prevent scattering into intermediate
states with a nucleon in any of the states defined by the orbitals
from the $1s_{1/2}$ to the $0g_{9/2}$ states or with
two nucleons in the $sdg$-shell ($0g_{9/2}$ excluded) or
two nucleons in the $pfh$-shell.
A harmonic-oscillator basis was chosen for the single-particle
wave functions, with an oscillator energy $\hbar\omega$, $\omega$ being
the oscillator frequency.

The shell model calculation for the Sn isotopes are based on
eq.\ (\ref{eq:eff}).
It is important for the present work to include all neutron configurations
in the model space. This means diagonalization of large
energy matrices. We have chosen a procedure based on the Lanczos algoritm
ref.~\cite{wwcm} where the states are given in the m-scheme Slater
determinant basis. The dimensionality of the matrices for different isotopes
are shown in table~\ref{tab:dim}.
This technique is an iterative procedure where the lowest states are
first calculated. At present $^{108}$Sn is not completed. Only
the four lowest states are found (see fig.~\ref{fig:2}).
\begin{table}[hbtp]
\caption{Dimension of the basis set in calculations of $^{102-108}$Sn.}
\label{tab:dim}
\begin{tabular}{l ccccccc}
\hline&&&&&&&\\
Isotope & $^{102}$Sn & $^{103}$Sn & $^{104}$Sn & $^{105}$Sn & $^{106}$Sn
& $^{107}$Sn  & $^{108}$Sn \\
\hline
Dimension & 36 & 245 & 1 504 & 7 451 & 31 124 & 108 297 & 323 682
\\&&&&&&&\\
\hline
\end{tabular}
\end{table}

\section{Results and discussion}
At present rather few energy levels are known for the lightest
Sn isotopes, a fact which makes a detailed comparison  difficult.
However, some characteristic properties  can be found in the data.
In the even isotopes there is a rather constant spacing between
the $0^{+}$ ground state and the first excited $2^{+}$ state.
Furthermore, in the odd systems a similar feature
is found between the $5/2^{+}$ ground state and the first excited
$7/2^{+}$ state up to $^{109}$Sn. In $^{111}$Sn these levels are
interchanged, with the $7/2^{+}$ state as the ground state.
This is well  described in the BCS model originally
analysed by Kisslinger and Sorensen~\cite{ks60} and also in the generalized
seniority scheme discussed by Talmi~\cite{tal77}.

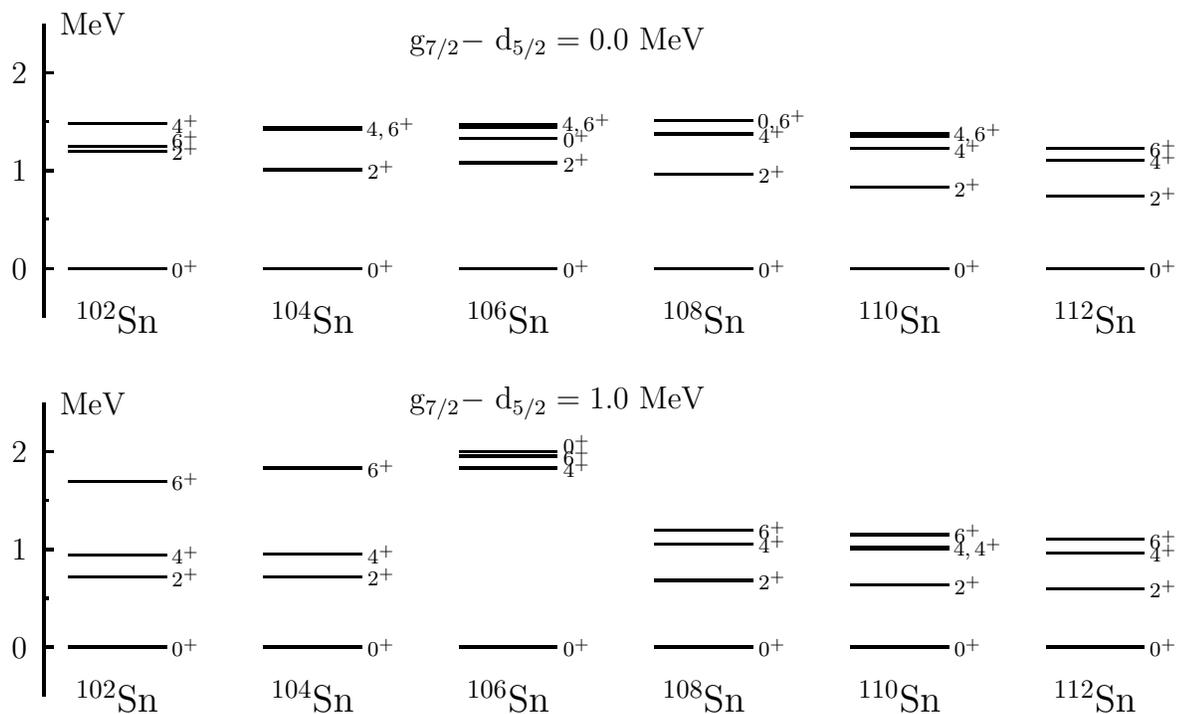
\begin{figure}[htbp]
\setlength{\unitlength}{1.3cm}

\begin{center}
\begin{picture}(8,3.5)(1,-1.5)

\newcommand{\lc}[1]{\put(-.5,#1){\line(1,0){1}}}
\newcommand{\ls}[2]{\put(.7,#1){\makebox(0,0){{\scriptsize $#2^{+}$}}}}
\newcommand{\lsr}[2]{\put(.9,#1){\makebox(0,0){{\scriptsize $#2^{+}$}}}}

\newcommand{\lcc}[1]{\put(1.5,#1){\line(1,0){1}}}
\newcommand{\lss}[2]{\put(2.7,#1){\makebox(0,0){{\scriptsize $#2^{+}$}}}}
\newcommand{\lssr}[2]{\put(2.8,#1){\makebox(0,0){{\scriptsize $#2^{+}$}}}}

\newcommand{\lccc}[1]{\put(3.5,#1){\line(1,0){1}}}
\newcommand{\lsss}[2]{\put(4.7,#1){\makebox(0,0){{\scriptsize $#2^{+}$}}}}
\newcommand{\lsssr}[2]{\put(4.8,#1){\makebox(0,0){{\scriptsize $#2^{+}$}}}}

\newcommand{\lcccc}[1]{\put(5.5,#1){\line(1,0){1}}}
\newcommand{\lssss}[2]{\put(6.7,#1){\makebox(0,0){{\scriptsize $#2^{+}$}}}}
\newcommand{\lssssr}[2]{\put(6.8,#1){\makebox(0,0){{\scriptsize $#2^{+}$}}}}

\newcommand{\lccccc}[1]{\put(7.5,#1){\line(1,0){1}}}
\newcommand{\lsssss}[2]{\put(8.7,#1){\makebox(0,0){{\scriptsize $#2^{+}$}}}}
\newcommand{\lsssssr}[2]{\put(8.8,#1){\makebox(0,0){{\scriptsize
$#2^{+}$}}}}

\newcommand{\lcccccc}[1]{\put(9.5,#1){\line(1,0){1}}}
\newcommand{\lssssss}[2]{\put(10.7,#1){\makebox(0,0){{\scriptsize
$#2^{+}$}}}}
\newcommand{\lssssssr}[2]{\put(10.8,#1){\makebox(0,0){{\scriptsize
$#2^{+}$}}}}

\put(-.25,2.5){\makebox(0,0){MeV}}
\put(4.5,2.3){\makebox(0,0){{g$_{7/2} -$ d$_{5/2} = 0.0$ MeV}}}

\thicklines
\put(-.75,-.5){\line(0,1){3}}
\multiput(-.75,.0)(0,1){3}{\line(1,0){.1}}
\multiput(-.75,.5)(0,1){2}{\line(1,0){.05}}

\put(-1.,2){\makebox(0,0){2}}
\put(-1.,1){\makebox(0,0){1}}
\put(-1.,0){\makebox(0,0){0}}


\lc{.0}      \ls{.0}{0}
\lc{1.194}   \ls{1.194}{2}
\lc{1.249}   \ls{1.319}{6}
\lc{1.478}   \ls{1.478}{4}
\put(0,-.5){\makebox(0,0){{\large $^{102}$Sn}}}

\lcc{.0}      \lss{.0}{0}
\lcc{1.009}   \lss{1.009}{2}
\lcc{1.423}   \lssr{1.423}{4,6}
\lcc{1.438}   
\put(2,-.5){\makebox(0,0){{\large $^{104}$Sn}}}

\lccc{.0}      \lsss{.0}{0}
\lccc{1.080}   \lsss{1.080}{2}
\lccc{1.332}   \lsss{1.332}{0}
\lccc{1.443}   \lsssr{1.473}{4,6}
\lccc{1.467}   
\put(4,-.5){\makebox(0,0){{\large $^{106}$Sn}}}

\lcccc{.0}      \lssss{.0}{0}
\lcccc{.963}    \lssss{.963}{2}
\lcccc{1.373}   \lssss{1.373}{4}
\lcccc{1.510}   \lssssr{1.510}{0,6}
\lcccc{1.515}   
\put(6,-.5){\makebox(0,0){{\large $^{108}$Sn}}}

\lccccc{.0}     \lsssss{.0}{0}
\lccccc{.829}   \lsssss{.829}{2}
\lccccc{1.225}  \lsssss{1.225}{4}
\lccccc{1.352}  \lsssssr{1.372}{4,6}
\lccccc{1.375}  
\put(8,-.5){\makebox(0,0){{\large $^{110}$Sn}}}

\lcccccc{.0}     \lssssss{.0}{0}
\lcccccc{.736}   \lssssss{.736}{2}
\lcccccc{1.104}  \lssssss{1.104}{4}
\lcccccc{1.231}  \lssssss{1.231}{6}
\put(10,-.5){\makebox(0,0){{\large $^{112}$Sn}}}

\end{picture}
\end{center}

\begin{center}
\begin{picture}(8,2)(1,0)

\newcommand{\lc}[1]{\put(-.5,#1){\line(1,0){1}}}
\newcommand{\ls}[2]{\put(.7,#1){\makebox(0,0){{\scriptsize $#2^{+}$}}}}
\newcommand{\lsr}[2]{\put(.8,#1){\makebox(0,0){{\scriptsize $#2^{+}$}}}}

\newcommand{\lcc}[1]{\put(1.5,#1){\line(1,0){1}}}
\newcommand{\lss}[2]{\put(2.7,#1){\makebox(0,0){{\scriptsize $#2^{+}$}}}}
\newcommand{\lssr}[2]{\put(2.8,#1){\makebox(0,0){{\scriptsize $#2^{+}$}}}}

\newcommand{\lccc}[1]{\put(3.5,#1){\line(1,0){1}}}
\newcommand{\lsss}[2]{\put(4.7,#1){\makebox(0,0){{\scriptsize $#2^{+}$}}}}
\newcommand{\lsssr}[2]{\put(4.8,#1){\makebox(0,0){{\scriptsize $#2^{+}$}}}}

\newcommand{\lcccc}[1]{\put(5.5,#1){\line(1,0){1}}}
\newcommand{\lssss}[2]{\put(6.7,#1){\makebox(0,0){{\scriptsize $#2^{+}$}}}}
\newcommand{\lssssr}[2]{\put(6.8,#1){\makebox(0,0){{\scriptsize $#2^{+}$}}}}

\newcommand{\lccccc}[1]{\put(7.5,#1){\line(1,0){1}}}
\newcommand{\lsssss}[2]{\put(8.7,#1){\makebox(0,0){{\scriptsize $#2^{+}$}}}}
\newcommand{\lsssssr}[2]{\put(8.8,#1){\makebox(0,0){{\scriptsize
$#2^{+}$}}}}

\newcommand{\lcccccc}[1]{\put(9.5,#1){\line(1,0){1}}}
\newcommand{\lssssss}[2]{\put(10.7,#1){\makebox(0,0){{\scriptsize
$#2^{+}$}}}}
\newcommand{\lssssssr}[2]{\put(10.8,#1){\makebox(0,0){{\scriptsize
$#2^{+}$}}}}

\put(-.25,2.5){\makebox(0,0){MeV}}
\put(4.5,2.5){\makebox(0,0){{g$_{7/2} - $ d$_{5/2} = 1.0$ MeV}}}

\thicklines
\put(-.75,-.5){\line(0,1){3}}
\multiput(-.75,.0)(0,1){3}{\line(1,0){.1}}
\multiput(-.75,.5)(0,1){2}{\line(1,0){.05}}

\put(-1.,2){\makebox(0,0){2}}
\put(-1.,1){\makebox(0,0){1}}
\put(-1.,0){\makebox(0,0){0}}


\lc{.0}      \ls{.0}{0}
\lc{.715}    \ls{.715}{2}
\lc{.944}    \ls{.944}{4}
\lc{1.700}   \ls{1.700}{6}
\put(0,-.5){\makebox(0,0){{\large $^{102}$Sn}}}

\lcc{.0}      \lss{.0}{0}
\lcc{.722}    \lss{.722}{2}
\lcc{.949}    \lss{.949}{4}
\lcc{1.832}   \lss{1.832}{6}
\put(2,-.5){\makebox(0,0){{\large $^{104}$Sn}}}

\lccc{.0}      \lsss{.0}{0}
\lccc{1.834}   \lsss{1.834}{4}
\lccc{1.958}   \lsss{1.958}{6}
\lccc{2.002}   \lsss{2.082}{0}
\put(4,-.5){\makebox(0,0){{\large $^{106}$Sn}}}

\lcccc{.0}      \lssss{.0}{0}
\lcccc{.682}    \lssss{.682}{2}
\lcccc{1.056}   \lssss{1.056}{4}
\lcccc{1.199}   \lssss{1.199}{6}
\put(6,-.5){\makebox(0,0){{\large $^{108}$Sn}}}

\lccccc{.0}     \lsssss{.0}{0}
\lccccc{.636}   \lsssss{.636}{2}
\lccccc{1.006}  \lsssssr{1.006}{4,4}
\lccccc{1.029}  
\lccccc{1.151}  \lsssss{1.161}{6}
\put(8,-.5){\makebox(0,0){{\large $^{110}$Sn}}}

\lcccccc{.0}     \lssssss{.0}{0}
\lcccccc{.598}   \lssssss{.598}{2}
\lcccccc{.965}   \lssssss{.965}{4}
\lcccccc{1.103}  \lssssss{1.103}{6}
\put(10,-.5){\makebox(0,0){{\large $^{112}$Sn}}}

\end{picture}
\end{center}

\caption{Energy spectra of $^{102-112}$Sn with a model space consisting
of the $g_{7/2}$ and $d_{5/2}$ single-particle orbitals.}
\label{fig:1}
\end{figure}

In our extended shell model calculation two elements are important
for the theoretical energy spectrum -- the single-particle
spectrum and the effective two-body matrix elements. We are reasonably
confident about the calculation of the effective force.
Similar methods have been used in the oxygen  and calcium
regions with reasonable results. However, the details
of the single-particle spectrum are rather uncertain.
In order to get a better understanding of the role of the single-particle
energies we have investigated a simplified model. The model space
is reduced to  the $g_{7/2}$ and $d_{5/2}$ orbitals only, and the relative
spacing has been varied. The results of two extreme cases, one with the
two orbitals degenerated and one with the two orbitals separated by 1.0 MeV,
are displayed in fig.~\ref{fig:1}. With this reduced model space the energy
spectra for the sequence of even Sn isotopes with mass number A = 102 -- 112
are calculated.
When the $g_{7/2}$ and $d_{5/2}$ orbitals become too strongly separated, the
$0^{+}$ -- $2^{+}$ spacing does not remain constant. The energy gap
increases
dramatically for $^{106}$Sn, where the $d_{5/2}$ orbital is nearly
filled.
Thus the nearly degenerate values for the single-particle energies
employed in ref.~\cite{grawe92} seem reasonable and are used in our
calculation. The calculated values of the higher single-particle levels have
been used without further discussion.

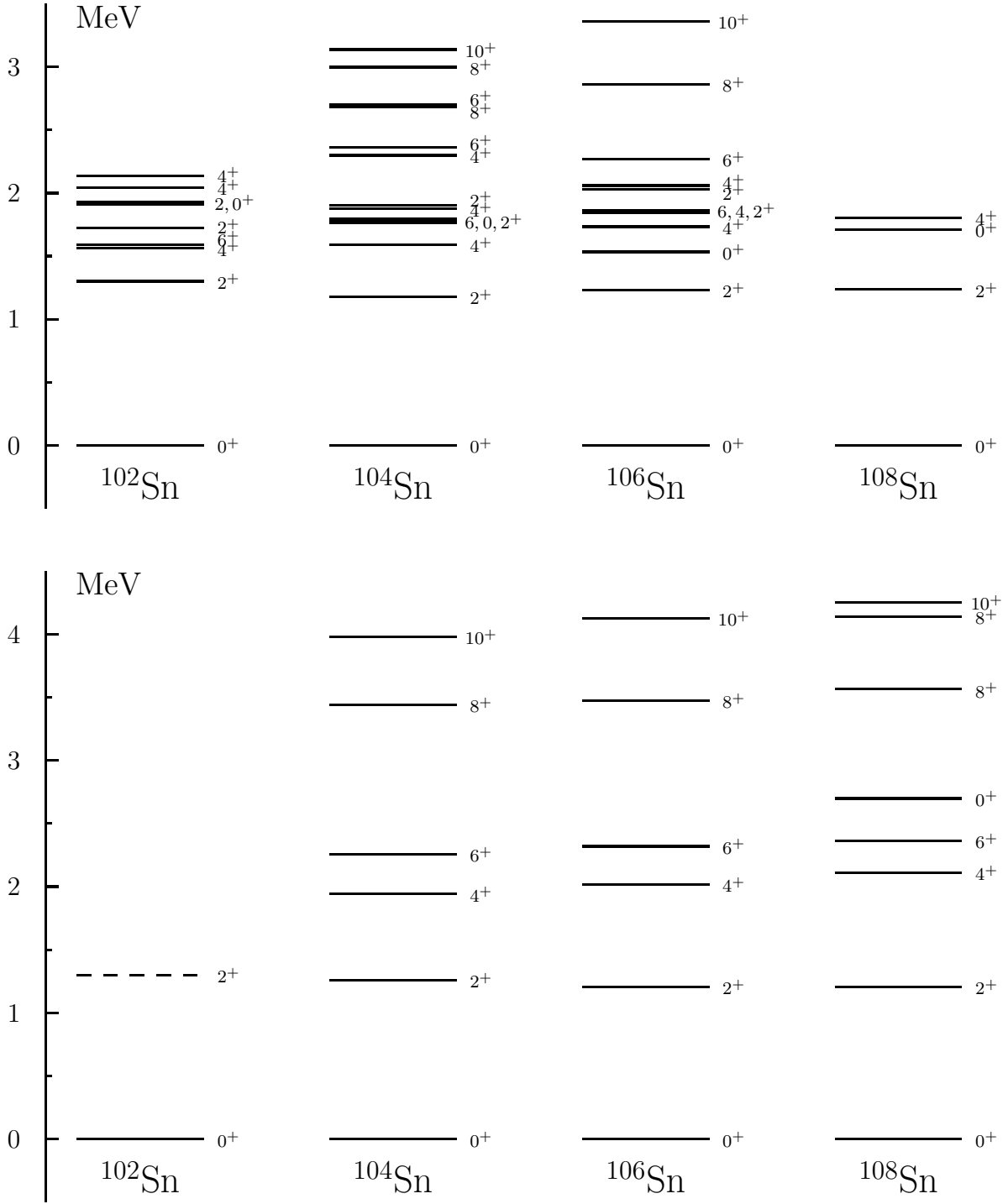
\begin{figure}[htbp]
\setlength{\unitlength}{2cm}
\begin{center}
\begin{picture}(4.9,4.5)(.05,-1.)
\newcommand{\lc}[1]{\put(-.5,#1){\line(1,0){1}}}
\newcommand{\ls}[2]{\put(.7,#1){\makebox(0,0){{\scriptsize $#2^{+}$}}}}
\newcommand{\lsr}[2]{\put(.75,#1){\makebox(0,0){{\scriptsize $#2^{+}$}}}}

\newcommand{\lcc}[1]{\put(1.5,#1){\line(1,0){1}}}
\newcommand{\lss}[2]{\put(2.7,#1){\makebox(0,0){{\scriptsize $#2^{+}$}}}}
\newcommand{\lssr}[2]{\put(2.8,#1){\makebox(0,0){{\scriptsize $#2^{+}$}}}}

\newcommand{\lccc}[1]{\put(3.5,#1){\line(1,0){1}}}
\newcommand{\lsss}[2]{\put(4.7,#1){\makebox(0,0){{\scriptsize $#2^{+}$}}}}
\newcommand{\lsssr}[2]{\put(4.8,#1){\makebox(0,0){{\scriptsize $#2^{+}$}}}}

\newcommand{\lcccc}[1]{\put(5.5,#1){\line(1,0){1}}}
\newcommand{\lssss}[2]{\put(6.7,#1){\makebox(0,0){{\scriptsize $#2^{+}$}}}}
\newcommand{\lssssr}[2]{\put(6.9,#1){\makebox(0,0){{\scriptsize $#2^{+}$}}}}

\newcommand{\lccccc}[1]{\put(7.5,#1){\line(1,0){1}}}
\newcommand{\lsssss}[2]{\put(8.7,#1){\makebox(0,0){{\scriptsize $#2^{+}$}}}}
\newcommand{\lsssssr}[2]{\put(8.9,#1){\makebox(0,0){{\scriptsize
$#2^{+}$}}}}

\put(-.25,3.4){\makebox(0,0){\large MeV}}

\thicklines
\put(-.75,-.5){\line(0,1){4}}
\multiput(-.75,.0)(0,1){4}{\line(1,0){.1}}
\multiput(-.75,.5)(0,1){3}{\line(1,0){.05}}

\put(-1.,3){\makebox(0,0){3}}
\put(-1.,2){\makebox(0,0){2}}
\put(-1.,1){\makebox(0,0){1}}
\put(-1.,0){\makebox(0,0){0}}

\lc{.0}      \ls{.0}{0}
\lc{1.302}   \ls{1.302}{2}
\lc{1.562}   \ls{1.562}{4}
\lc{1.592}   \ls{1.64}{6}
\lc{1.724}   \ls{1.740}{2}
\lc{1.911}   \lsr{1.911}{2,0}
\lc{1.925}   
\lc{2.045}   \ls{2.045}{4}
\lc{2.137}   \ls{2.147}{4}
\put(0,-.3){\makebox(0,0){{\Large $^{102}$Sn}}}

\lcc{.0}      \lss{.0}{0}
\lcc{1.179}   \lss{1.179}{2}
\lcc{1.589}   \lss{1.589}{4}
\lcc{1.763}   \lssr{1.763}{6,0,2}
\lcc{1.780}   
\lcc{1.795}   
\lcc{1.876}   \lss{1.870}{4}
\lcc{1.903}   \lss{1.953}{2}
\lcc{2.299}   \lss{2.299}{4}
\lcc{2.361}   \lss{2.400}{6}
\lcc{2.683}   \lss{2.653}{8}
\lcc{2.700}   \lss{2.750}{6}
\lcc{2.996}   \lss{2.996}{8}
\lcc{3.137}   \lss{3.137}{10}

\put(2,-.3){\makebox(0,0){{\Large $^{104}$Sn}}}

\lccc{.0}      \lsss{.0}{0}
\lccc{1.233}   \lsss{1.233}{2}
\lccc{1.535}   \lsss{1.535}{0}
\lccc{1.732}   \lsss{1.732}{4}
\lccc{1.843}   
\lccc{1.849}   \lsssr{1.849}{6,4,2}
\lccc{1.860}   
\lccc{2.027}   \lsss{2.007}{2}
\lccc{2.058}   \lsss{2.090}{4}
\lccc{2.269}   \lsss{2.269}{6}
\lccc{2.862}   \lsss{2.862}{8}
\lccc{3.361}   \lsss{3.361}{10}
\put(4,-.3){\makebox(0,0){{\Large $^{106}$Sn}}}

\lcccc{.0}    \lssss{.0}{0}
\lcccc{1.236} \lssss{1.231}{2}
\lcccc{1.710} \lssss{1.710}{0}
\lcccc{1.803} \lssss{1.803}{4}
\put(6,-.3){\makebox(0,0){{\Large $^{108}$Sn}}}
\end{picture}
\end{center}

\begin{center}
\begin{picture}(4.9,4.5)(.05,-.25)

\newcommand{\lc}[1]{\put(-.5,#1){\line(1,0){1}}}
\newcommand{\ls}[2]{\put(.7,#1){\makebox(0,0){{\scriptsize $#2^{+}$}}}}
\newcommand{\lsr}[2]{\put(.9,#1){\makebox(0,0){{\scriptsize $#2^{+}$}}}}

\newcommand{\lcc}[1]{\put(1.5,#1){\line(1,0){1}}}
\newcommand{\lss}[2]{\put(2.7,#1){\makebox(0,0){{\scriptsize $#2^{+}$}}}}
\newcommand{\lssr}[2]{\put(2.9,#1){\makebox(0,0){{\scriptsize $#2^{+}$}}}}

\newcommand{\lccc}[1]{\put(3.5,#1){\line(1,0){1}}}
\newcommand{\lsss}[2]{\put(4.7,#1){\makebox(0,0){{\scriptsize $#2^{+}$}}}}
\newcommand{\lsssr}[2]{\put(4.9,#1){\makebox(0,0){{\scriptsize $#2^{+}$}}}}

\newcommand{\lcccc}[1]{\put(5.5,#1){\line(1,0){1}}}
\newcommand{\lssss}[2]{\put(6.7,#1){\makebox(0,0){{\scriptsize $#2^{+}$}}}}
\newcommand{\lssssr}[2]{\put(6.9,#1){\makebox(0,0){{\scriptsize $#2^{+}$}}}}

\put(-.25,4.4){\makebox(0,0){\large MeV}}

\thicklines
\put(-.75,-.5){\line(0,1){5}}
\multiput(-.75,.0)(0,1){5}{\line(1,0){.1}}
\multiput(-.75,.5)(0,1){4}{\line(1,0){.05}}

\put(-1.,4){\makebox(0,0){4}}
\put(-1.,3){\makebox(0,0){3}}
\put(-1.,2){\makebox(0,0){2}}
\put(-1.,1){\makebox(0,0){1}}
\put(-1.,0){\makebox(0,0){0}}

\lc{.0}       \ls{.0}{0}
\put(-.5,1.3){\line(1,0){.11}}
\put(-.29,1.3){\line(1,0){.11}}
\put(-.08,1.3){\line(1,0){.11}}
\put(.13,1.3){\line(1,0){.11}}
\put(.34,1.3){\line(1,0){.11}}
                \ls{1.3}{2}

\put(0,-.3){\makebox(0,0){{\Large $^{102}$Sn}}}

\lcc{.0}      \lss{.0}{0}
\lcc{1.259}   \lss{1.259}{2}
\lcc{1.942}   \lss{1.942}{4}
\lcc{2.257}   \lss{2.257}{6}
\lcc{3.440}   \lss{3.440}{8}
\lcc{3.980}   \lss{3.980}{10}

\put(2,-.3){\makebox(0,0){{\Large $^{104}$Sn}}}

\lccc{.0}      \lsss{.0}{0}
\lccc{1.207}   \lsss{1.207}{2}
\lccc{2.017}   \lsss{2.017}{4}
\lccc{2.321}   \lsss{2.321}{6}
\lccc{3.476}   \lsss{3.476}{8}
\lccc{4.128}   \lsss{4.128}{10}
\put(4,-.3){\makebox(0,0){{\Large $^{106}$Sn}}}

\lcccc{.0}      \lssss{.0}{0}
\lcccc{1.206}    \lssss{1.206}{2}
\lcccc{2.112}   \lssss{2.112}{4}
\lcccc{2.365}   \lssss{2.365}{6}
\lcccc{2.700}   \lssss{2.700}{0}
\lcccc{3.566}   \lssss{3.556}{8}
\lcccc{4.140}   \lssss{4.140}{8}
\lcccc{4.251}   \lssss{4.251}{10}
\put(6,-.3){\makebox(0,0){{\Large $^{108}$Sn}}}
\end{picture}
\end{center}
\caption{Theoretical (upper) and experimental (lower) energy spectra
for the even isotopes $^{102-108}$Sn. The experimental $2^{+}$ level
(dashed) in $^{102}$Sn is tentative [19].}
\label{fig:2}
\end{figure}

\begin{figure}[htbp]
\setlength{\unitlength}{4cm}
\begin{center}
\begin{picture}(2.92,2.5)(-.4,-.25)
\newcommand{\lc}[1]{\put(-.5,#1){\line(1,0){.5}}}
\newcommand{\ls}[2]{\put(.15,#1){\makebox(0,0){{\scriptsize $#2^{+}$}}}}
\newcommand{\lsr}[2]{\put(.20,#1){\makebox(0,0){{\scriptsize $#2^{+}$}}}}

\newcommand{\lcc}[1]{\put(.8,#1){\line(1,0){.5}}}
\newcommand{\lss}[2]{\put(1.45,#1){\makebox(0,0){{\scriptsize $#2^{+}$}}}}
\newcommand{\lssr}[2]{\put(1.50,#1){\makebox(0,0){{\scriptsize $#2^{+}$}}}}

\newcommand{\lccc}[1]{\put(2.1,#1){\line(1,0){.5}}}
\newcommand{\lsss}[2]{\put(2.75,#1){\makebox(0,0){{\scriptsize $#2^{+}$}}}}
\newcommand{\lsssr}[2]{\put(2.80,#1){\makebox(0,0){{\scriptsize $#2^{+}$}}}}

\put(-.5,2.2){\makebox(0,0){\large MeV}}

\thicklines
\put(-.75,-.25){\line(0,1){2.5}}
\multiput(-.75,.0)(0,1){3}{\line(1,0){.05}}
\multiput(-.75,.5)(0,1){2}{\line(1,0){.025}}

\put(-.85,2){\makebox(0,0){2}}
\put(-.85,1){\makebox(0,0){1}}
\put(-.85,0){\makebox(0,0){0}}

\lc{0}        \ls{0}{5/2}
\lc{0.291}    \ls{.291}{7/2}
\lc{0.758}    \ls{.758}{3/2}
\lc{1.145}    \ls{1.145}{9/2}
\lc{1.210}    \ls{1.200}{5/2}
\lc{1.231}    \ls{1.231}{9/2}
\lc{1.293}    \lsr{1.293}{11/2,7/2}
\lc{1.297}    
\lc{1.332}    \ls{1.332}{7/2}
\lc{1.390}    \ls{1.390}{15/2}
\lc{1.607}    \ls{1.607}{13/2}
\lc{1.797}    \ls{1.797}{17/2}

\put(-.25,-.15){\makebox(0,0){{\Large $^{103}$Sn}}}

\lcc{0}    \lss{0}{5/2}
\lcc{0.299}    \lss{.299}{7/2}
\lcc{0.781}    \lss{.781}{3/2}
\lcc{1.084}    \lssr{1.084}{3/2,5/2}
\lcc{1.096}    
\lcc{1.154}    \lss{1.145}{11/2}
\lcc{1.183}    \lss{1.199}{9/2}
\lcc{1.277}    \lss{1.277}{5/2}
\lcc{1.364}    \lss{1.360}{7/2}
\lcc{1.393}    \lss{1.420}{7/2}
\lcc{1.542}    \lss{1.542}{13/2}
\lcc{1.592}    \lss{1.592}{11/2}
\lcc{1.614}    \lss{1.640}{15/2}
\lcc{1.755}    \lss{1.755}{13/2}
\lcc{1.825}    \lss{1.825}{11/2}
\lcc{1.898}    \lss{1.898}{15/2}
\lcc{1.950}    \lss{1.950}{17/2}

\put(1.05,-.15){\makebox(0,0){{\Large $^{105}$Sn}}}

\lccc{0}    \lsss{0}{5/2}
\lccc{0.079}    \lsss{.079}{7/2}
\lccc{0.866}    \lsss{.866}{5/2}
\lccc{0.902}    \lsss{.932}{3/2}
\lccc{1.033}    \lsssr{1.033}{9/2,1/2}
\lccc{1.039}    
\lccc{1.137}    \lsssr{1.137}{3/2,7/2}
\lccc{1.143}    
\lccc{1.208}    \lsssr{1.208}{11/2,5/2}
\lccc{1.217}    
\lccc{1.241}    \lsss{1.271}{7/2}
\lccc{1.376}    \lsss{1.376}{9/2}
\lccc{1.418}    \lsss{1.438}{3/2}
\lccc{1.515}    \lsss{1.515}{13/2}
\lccc{1.549}    \lsss{1.579}{11/2}
\lccc{1.686}    \lsss{1.676}{15/2}
\lccc{1.748}    
\lccc{1.756}    \lsssr{1.766}{13/2,15/2}
\lccc{1.714}    \lsss{1.724}{11/2}
\lccc{1.821}    \lsss{1.821}{17/2}


\put(2.35,-.15){\makebox(0,0){{\Large $^{107}$Sn}}}

\end{picture}
\end{center}

\begin{center}
\begin{picture}(2.92,2.5)(-.4,-.15)

\newcommand{\lc}[1]{\put(-.5,#1){\line(1,0){.5}}}
\newcommand{\ls}[2]{\put(.15,#1){\makebox(0,0){{\scriptsize $#2^{+}$}}}}
\newcommand{\lsr}[2]{\put(.35,#1){\makebox(0,0){{\scriptsize $#2^{+}$}}}}

\newcommand{\lcc}[1]{\put(.8,#1){\line(1,0){.5}}}
\newcommand{\lss}[2]{\put(1.45,#1){\makebox(0,0){{\scriptsize $#2^{+}$}}}}
\newcommand{\lssr}[2]{\put(1.65,#1){\makebox(0,0){{\scriptsize $#2^{+}$}}}}

\newcommand{\lccc}[1]{\put(2.1,#1){\line(1,0){.5}}}
\newcommand{\lsss}[2]{\put(2.75,#1){\makebox(0,0){{\scriptsize $#2^{+}$}}}}
\newcommand{\lsssr}[2]{\put(2.95,#1){\makebox(0,0){{\scriptsize $#2^{+}$}}}}

\put(-.5,2.2){\makebox(0,0){\large MeV}}

\thicklines
\put(-.75,-.25){\line(0,1){2.5}}
\multiput(-.75,.0)(0,1){3}{\line(1,0){.05}}
\multiput(-.75,.5)(0,1){2}{\line(1,0){.025}}

\put(-.85,2){\makebox(0,0){2}}
\put(-.85,1){\makebox(0,0){1}}
\put(-.85,0){\makebox(0,0){0}}

\lcc{0.0}        \lss{0.0}{5/2}
\lcc{0.200}      \lss{.200}{7/2}
\lcc{1.195}      \lss{1.195}{9/2}
\lcc{1.393}      \lss{1.393}{11/2}
\lcc{1.849}      \lss{1.849}{13/2}
\lcc{2.031}      \lss{2.031}{15/2}
\lcc{2.204}      \lss{2.204}{17/2}

\put(1.05,-.15){\makebox(0,0){{\Large $^{105}$Sn}}}

\lccc{0.0}        \lsss{0.0}{5/2}
\lccc{0.151}      \lsss{.151}{7/2}
\lccc{1.221}      \lsss{1.221}{9/2}
\lccc{1.350}      \lsss{1.340}{11/2}
\lccc{1.371}      \lsss{1.391}{9/2}
\lccc{1.798}      \lsss{1.798}{13/2}
\lccc{1.943}      \lsss{1.943}{13/2}
\lccc{2.067}      \lsss{2.067}{15/2}

\put(2.35,-.15){\makebox(0,0){{\Large $^{107}$Sn}}}
\end{picture}
\end{center}
\caption{Theoretical (upper) and experimental (lower) energy spectra
for the odd isotopes $^{103-107}$Sn.}
\label{fig:3}
\end{figure}
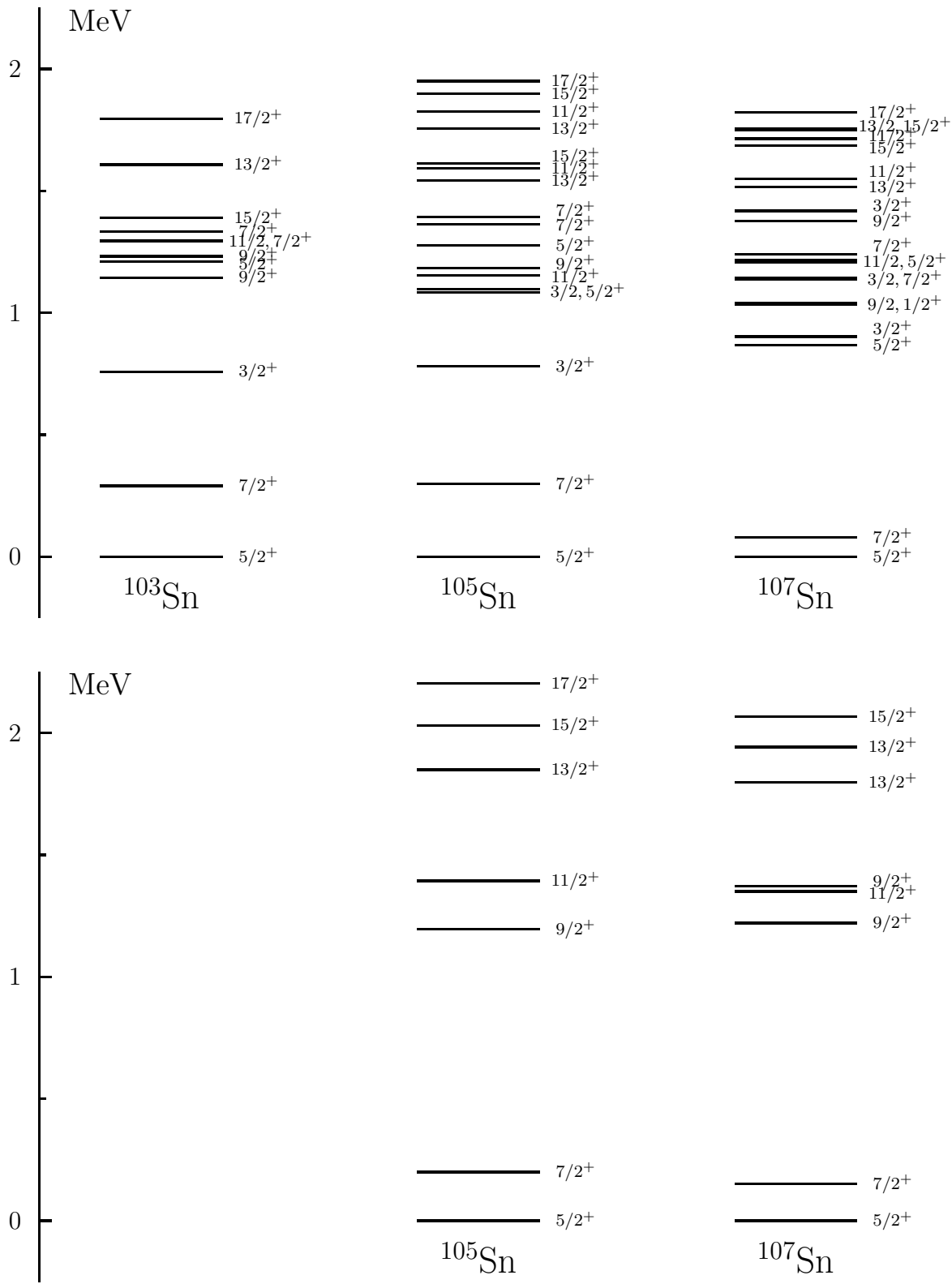

The resulting energy spectra obtained with a complete single-particle basis
including the $h_{11/2}$ orbital and the above-mentioned
effective interaction are displayed in fig.~\ref{fig:2} and
fig.~\ref{fig:3}.
For the even Sn isotopes (fig.~\ref{fig:2}) the excitation energies
of the lowest $J=2$ states  are in good agreement with
the experimental values, with a spacing of about 1.2 MeV.
The theoretical spectra above the first excited $J = 2$ state
are far more difficult to interpret.
Compared to the levels experimentally found our calculation gives many
more levels in the region between 1 and 2~MeV. However, we believe
that there are still more experimental levels to be found
in this region in the future.

The odd Sn isotopes are displayed in fig.~\ref{fig:3}. The two lowest energy
levels $J=5/2^{+}$ and $J=7/2^{+}$ are reproduced in correct order, but
in $^{107}$Sn the theoretical spacing is smaller than the experimental
value. This may in turn lead to a
$J=7/2^{+}$ ground state in $^{109}$Sn, contrary to what
is experimentally found.
Moreover, this
could indicate that the higher single-particle levels should
be lowered in order to reduce the filling of the $d_{5/2}$ and the
$g_{7/2}$
levels. As in the even isotopes many more energy levels are calculated
than known experimentally.

A further test of our results and the properties of the model wave
function may be found in the two known
$6^{+} \rightarrow 4^{+}$ E2 transitions~\cite{grawe92}.
There are two measured E2 transitions,
one in $^{104}$Sn $6^{+} \rightarrow 4^{+}$ with a strength of 4.0 Wu and
one in $^{104}$Sn $6^{+} \rightarrow 4^{+}$ with strength 2.5 Wu.
With the assumption that the observed $4^{+}$ and $6^{+}$ states
are the lowest ones, our model is not able to reproduce
the E2 transitions with sufficient
strength. An alternative interpretation may be that the E2 transitions
observed are between higher-lying states. We have calculated
two $6^{+} \rightarrow 4^{+}$ E2 transitions with strengths 2.1 Wu and
3.0 Wu, which are comparable to the observations in $^{104}$Sn and
$^{106}$Sn, respectively. The transitions are calculated with an effective
neutron charge $e_{n} = 1.0e$, a value which is within the range of
theoretically calculated effective charges for the Sn isotopes
\cite{ehho94}.
Based on these values we may interpret  our results in the following way:
In $^{104}$Sn the $4^{+}$ state at 1.942 MeV corresponds to our $4^{+}_{2}$
state and the $6^{+}$ state at 2.257 MeV to our $6^{+}_{2}$ state.
Similarily
in $^{106}$Sn the $4^{+}$ state at 2.017 MeV corresponds to our $4^{+}_{3}$
state and the $6^{+}$ state at 2.321 MeV to our $6^{+}_{2}$ state.

\clearpage
\clearpage



\begin{thebibliography}{99}
\bibitem{schu92}  R. Schubart {\em et al.}, Z. Phys. {\bf A343} (1992) 123
\bibitem{schu91}  R. Schubart {\em et al.}, Z. Phys. {\bf A340} (1991) 109
\bibitem{grawe92}  H. Grawe {\em et al.}, Prog. Part. Nucl. Phys. {\bf 28}
(1992) 281
\bibitem{ryk92}  A. Plochocki {\em et al.}, Z. Phys. {\bf A342} (1992) 43
\bibitem{sw86}  S.D. Serot and J.D. Walecka, Adv. Nucl. Phys. {\bf 16}
(1986)
1
\bibitem{hir91}  D. Hirata {\em et al.}, Phys. Rev. {\bf C 44}, (1991) 1467
\bibitem{nhm92}  T. Nikolaus, T. Hoch and D.G. Madland,
Phys. Rev. {\bf C~46} (1992) 1757
\bibitem{lean84}  G.A. Leander, J. Dudek, W. Nazarewicz, J.R. Nix and Ph.
Quentin, Phys. Rev. {\bf C 30} (1984) 416
\bibitem{tal77} I. Talmi in Elementary Modes of Excitation in Nuclei,
ed. A. Bohr and R.A. Broglia, (North-Holland, Amsterdam, 1977);
private communication
\bibitem{mfko83} H. M\"{u}ther, A. Faessler, T.T.S. Kuo and E. Osnes,
Nucl.\ Phys. {\bf 401} (1983) 124;
H. M\"{u}ther, A. Polls and T.T.S. Kuo, Nucl.\ Phys. {\bf 435} (1985) 548
\bibitem{ehho93}  T. Engeland, M. Hjorth-Jensen, A. Holt and E. Osnes,
Phys. Rev. {\bf C 48} (1993) 535
\bibitem{ko90}  T.T.S. Kuo and E. Osnes, Folded-Diagram Theory of the
Effective Interaction in Atomic Nuclei, Springer Lecture Notes in Physics,
(Springer, Berlin, 1990) Vol. 364
\bibitem{mac89}  R. Machleidt, Adv. Nucl. Phys. {\bf 19} (1989) 189
\bibitem{ls80} K.\ Suzuki and S.Y.\ Lee, Prog.\ Theor.\ Phys.\ {\bf 64}
(1980) 2091
\bibitem{hom92}  M. Hjorth-Jensen, E. Osnes and H. M\"uther, Ann. of Phys.
{\bf 213} (1992) 102
Nucl.
\bibitem{kkko76}  E.M. Krenciglowa, C.L. Kung, T.T.S. Kuo and E. Osnes, Ann.
of Phys. {\bf 101} (1976) 154
\bibitem{wwcm} R.R. Whitehead, A. Watt, B.J. Cole and I. Morrison,
Adv.\ Nucl.\ Phys.\ {\bf 7} (1977) 123
\bibitem{ks60} L.S. Kisslinger and R.A. Sorensen, Mat.\ Fys.\ Medd.\ Dan.\
Vid.\ Selsk.\ {\bf 32} (1960) 1
\bibitem{blom93} J. Blomqvist, private communication
\bibitem{ehho94} T. Engeland, M. Hjorth-Jensen, A. Holt and E. Osnes,
unpublished
\end{thebibliography}
\end{document}